# Beyond Chat: A Framework for LLMs as Human-Centered Support Systems

Zhiyin Zhou

New York, New York, USA

## Abstract

*Large language models are moving beyond transactional question answering to act as companions, coaches, mediators, and curators that scaffold human growth, decision-making, and well-being. This paper proposes a role-based framework for human-centered LLM support systems, compares real deployments across domains, and identifies cross-cutting design principles: transparency, personalization, guardrails, memory with privacy, and a balance of empathy and reliability. It outlines evaluation metrics that extend beyond accuracy to trust, engagement, and longitudinal outcomes. It also analyzes risks including over-reliance, hallucination, bias, privacy exposure, and unequal access, and proposes future directions spanning unified evaluation, hybrid human–AI models, memory architectures, cross-domain benchmarking, and governance. The goal is to support responsible integration of LLMs in sensitive settings where people need accompaniment and guidance, not only answers.*

## 1. Introduction

The rise of large language models (LLMs) has fundamentally reshaped the landscape of human–computer interaction. Tools such as OpenAI's GPT-4, Anthropic's Claude, and Google's Gemini have demonstrated unprecedented capabilities in natural language understanding, content generation, and conversational engagement. While early adoption has focused heavily on transactional uses such as drafting text, answering factual questions, or summarizing information, the potential of LLMs extends far beyond these narrow tasks [6], [7]. Increasingly, researchers and commercial platforms are exploring how LLMs might act as interactive partners that support not only productivity but also human growth, emotional well-being, and social inclusion [8], [9], [10], [11].

Despite these advances, most discussions of LLM utility continue to privilege accuracy, efficiency, and scale as primary evaluation criteria. Benchmarks in natural language processing typically emphasize factual correctness, linguistic fluency, and model robustness across diverse datasets [6]. Yet these criteria, while important, do not fully capture the role that LLMs are beginning to play in human lives. Decades ago, Weizenbaum warned that humans are prone to anthropomorphizing conversational systems [1], and Reeves and Nass showed that people treat media and computers as social actors [3]. These insights explain why millions of users now turn to LLM-based systems not simply for answers but for companionship, coaching, mediation, and knowledge curation. In such contexts, the central question is not only what the model produces but also how it scaffolds the user's cognitive, emotional, or social development [2], [5].

The growing reliance on LLMs for sensitive forms of support reflects broader social challenges. Populations worldwide face shortages of human resources in critical areas: the rising demand for mental health services vastly outpaces the availability of trained therapists [8], [9]; legal aid remains financially inaccessible for many [12]; educational inequalities persist despite the expansion of online learning [10], [11]; and aging societies are increasingly burdened by a lack of affordable caregivers [4]. Against this backdrop, LLMs present themselves as scalable, always-available, and low-cost technologies that can fill gaps where traditional human support is scarce or absent [6]. Importantly, however, this opportunity is coupled with significant risk. Unregulated or poorly designed systems may mislead users, foster over-reliance, or generate harm in precisely those domains where individuals are most vulnerable [7], [13], [17].

This paper argues that to responsibly integrate LLMs into sensitive human-support contexts, they must be reconceptualized not merely as conversational engines but as human-centered support systems. Drawing inspiration from education theory [2], psychotherapy, and human–computer interaction [5], [6], this paper proposes a comparative framework that captures four emerging roles of LLMs:

- **Companion**, providing emotional presence and mitigating loneliness [8], [9].
- **Coach**, guiding skill development through iterative feedback [10], [11].
- **Mediator**, bridging gaps in access such as legal interpretation or cross-lingual communication [12], [16].
- **Curator**, filtering overwhelming information into trusted and actionable insights [14], [15].

This framework provides a vocabulary for analyzing the diverse ways in which LLMs are already entering everyday life, while also identifying the design principles required to ensure that these systems remain trustworthy and supportive rather than manipulative or harmful [4], [6].

The contributions of this paper are fourfold. First, it consolidates scattered examples of LLM applications into a unifying conceptual framework that distinguishes between companion, coach, mediator, and curator roles. Second, it conducts a comparative analysis of real-world case studies ranging from language learning platforms to AI companionship tools to identify patterns and divergences in design requirements. Third, it articulates a set of evaluation metrics that move beyond accuracy, emphasizing trust, long-term engagement, and human growth outcomes [13], [17]. Finally, it highlights critical risks and outlines future directions for building hybrid human–AI support systems that combine the scalability of LLMs with the accountability of human oversight [4], [16].

By shifting the focus from transactional outputs to relational roles, this paper seeks to advance both theoretical understanding and practical design of LLMs in human-centered contexts. In doing so, it supports a more nuanced conversation about what it means to build artificial systems that not only answer but also accompany, guide, and empower the humans who rely on them.

This paper is conceptual in nature rather than empirical. Its contribution lies in synthesizing prior research and real-world deployments of large language models into a comparative framework that identifies four roles: companion, coach, mediator, and curator. The aim is not to present new experimental data but to establish a theoretical foundation, supported by illustrative case studies, that can guide future empirical research and system design [2], [4], [5]. By explicitly acknowledging this limitation, the paper positions itself as a starting point and invites future validation through user studies, domain-specific evaluations, and cross-cultural analyses.

## 2. BACKGROUND AND LITERATURE REVIEW

### 2.1. Evolution of Conversational Agents

The idea of machines capable of human-like conversation has fascinated computer scientists since the earliest days of artificial intelligence. One of the first notable systems was ELIZA, developed in 1966 by Joseph Weizenbaum [1], which simulated a Rogerian psychotherapist through scripted text transformations. Although ELIZA had no genuine understanding of meaning, many users attributed emotional depth to its simple responses. This early phenomenon foreshadowed a recurring challenge in conversational AI: people often project human qualities onto systems that are fundamentally rule-based [3].

Subsequent decades saw the emergence of commercial conversational agents such as Microsoft's Clippy, Apple's Siri, Amazon's Alexa, and Google Assistant. These systems combined natural language processing with information retrieval, enabling users to perform tasks such as setting reminders or searching the web. However, their conversational abilities remained limited to pre-programmed queries and domain-specific commands [6].

The advent of transformer-based architectures and large-scale pretraining on internet-scale corpora marked a turning point. LLMs such as GPT-3 and GPT-4 demonstrated not only the ability to generate coherent text but also emergent capabilities in reasoning, summarization, and style adaptation [7]. These advances made it possible to envision conversational systems that could provide nuanced, context-sensitive, and sustained interactions across multiple domains of human activity.

### 2.2. Applications of LLMs in Practice

The wide availability of LLMs has spurred a wave of applied research and commercial experimentation. Several domains illustrate both the promise and limitations of these systems.
In healthcare, LLMs are increasingly investigated for triage, decision support, and patient interaction. Social robotics projects such as ElliQ integrate conversational capabilities into companion robots designed to reduce loneliness among older adults [8].

In education, LLM-powered tutors are beginning to reshape learning. Khanmigo, developed in collaboration with OpenAI, provides a Socratic-style tutor that engages students in guided problem-solving [11]. Duolingo Max leverages GPT-4 to create role-play scenarios and interactive explanations for language learners [10].

In law and access to justice, systems such as DoNotPay attempt to democratize legal support by automating tasks such as contesting parking tickets or drafting appeals [12].

In research and knowledge discovery, tools such as Perplexity AI, Consensus, and Elicit position themselves as AI curators [14], [15]. These applications highlight the importance of verifiability, since users rely on the outputs for academic and professional work.

Together, these examples demonstrate that LLMs are not confined to single-use tasks. Instead, they are increasingly integrated into domains where human growth, decision-making, and well-being are central.

## 2.3. Human-Centered Theories Relevant to LLMs

Understanding the role of LLMs in sensitive human contexts requires more than technical evaluation. It also draws from theories in psychology, education, and human–computer interaction.

In educational theory, Lev Vygotsky's concept of the Zone of Proximal Development (ZPD) is particularly relevant [2]. ZPD describes the difference between what a learner can achieve independently and what they can achieve with guided assistance. In this context, LLMs may function as scaffolding agents, providing incremental support that helps learners progress beyond their current capacity.

In psychotherapy, the concept of a therapeutic alliance emphasizes the relationship between client and therapist as a key determinant of outcomes. Translated to AI, this suggests that the effectiveness of companion systems depends not only on what is said but on how the system fosters trust, empathy, and continuity over time. Studies of Replika users demonstrate that some individuals perceive emotional connection and comfort even when fully aware that the interaction is artificial [8], [9].

In human–computer interaction (HCI), researchers have emphasized the importance of transparency, explainability, and user trust [5], [6]. LLMs heighten these concerns because their generative nature can produce outputs that are convincing yet factually incorrect [7]. Designing for interpretability and verifiability is therefore crucial, especially when LLMs act as mediators or curators in high-stakes domains.

## 2.4. Gap Analysis

While the above literature demonstrates substantial progress in LLM applications, several gaps remain. First, the existing body of work tends to be siloed within specific domains such as healthcare, law, or education. This creates a fragmented picture of what LLMs can do, limiting the ability to generalize design principles across contexts.

Second, most evaluations rely on traditional metrics such as accuracy, latency, and user satisfaction [13]. These measures are necessary but insufficient when LLMs take on relational roles. For example, a tutoring system that answers questions correctly but discourages student confidence may succeed on accuracy but fail in its larger purpose.

Third, few studies attempt to compare the different ways in which LLMs scaffold human growth and support [17]. While there is scattered evidence that LLMs can act as companions, coaches, mediators, or curators, there is no unifying framework that organizes these roles or analyzes their distinct design requirements.

This paper addresses these gaps by developing a comparative framework for LLMs as human-centered support systems. The following section outlines the framework in detail, categorizing the roles of companion, coach, mediator, and curator, and highlighting the design principles necessary for their responsible development [4].

# 3. CONCEPTUAL FRAMEWORK: LLMs AS HUMAN-CENTERED SUPPORT SYSTEMS

## 3.1. Defining Human-Centered Support

Human-centered support refers to systems designed not only to provide correct information but also to foster growth, resilience, and empowerment in their users [6]. Unlike transactional tools that complete discrete tasks, human-centered systems engage in sustained interaction, adapt to evolving needs, and respond to the social and emotional dimensions of human experience. Large language models have begun to occupy this category, particularly as they are embedded in applications where users expect more than mere efficiency [7].

The distinction is important because traditional metrics of artificial intelligence such as accuracy, speed, and computational efficiency fail to capture the relational qualities that define support [5], [6]. A system that consistently provides factually accurate responses might still fail as a supportive companion if it cannot express empathy, or as an educational coach if it does not encourage persistence in the face of mistakes. Conversely, an imperfect system may still generate meaningful impact if it scaffolds confidence, engagement, or a sense of belonging [2]. Thus, the conceptual framework introduced here emphasizes the roles that LLMs assume in relation to human needs, rather than the technical mechanisms that underlie their outputs.

## 3.2. Four Roles of LLMs in Support Contexts

Through comparative analysis of existing applications and theoretical traditions, four primary roles are identified: companion, coach, mediator, and curator [4]. Each role reflects a distinct orientation toward the human user and comes with specific design requirements.

### 3.2.1. LLMs as Companions

The companion role involves providing emotional presence, empathy, and a sense of connection. Systems such as Replika and Character.AI exemplify this function by creating environments in which users can share feelings and receive supportive responses [8], [9]. In elder care, companion robots with conversational capabilities reduce isolation and help structure daily routines [8].

Key requirements for this role include emotional sensitivity, personalization to user mood and context, and boundaries that prevent over-dependence. Unlike transactional systems, companions succeed when they foster continuity, trust, and psychological comfort, even if their factual accuracy is limited [17]. The greatest risks arise when systems blur the line between artificial and human relationships or fail to maintain safe conversational norms [8], [9].

### 3.2.2. LLMs as Coaches

The coach role emphasizes guided skill development through structured feedback. Educational platforms such as Duolingo Max and Khanmigo illustrate this orientation by leveraging LLMs to generate practice scenarios, role-play exercises, and Socratic-style questioning that encourage learners to think critically rather than rely solely on answers [10], [11].

Coaching applications are not limited to academic domains. LLMs have also been explored in areas such as mental health support, where chatbots function as conversational coaches that help users reflect on emotions and practice coping strategies [15]. In addition, early studies have examined how LLMs can act as training mediators, providing structured simulations for dispute resolution and professional development [16]. These examples highlight the adaptability of the coaching paradigm across both educational and non-academic contexts.

Effective coaching requires transparency in feedback, adaptability to learner progress, and pacing that sustains motivation. The LLM must act as scaffolding, encouraging the user to reach beyond current abilities without overwhelming them. Risks emerge when the system provides inaccurate content, discouraging feedback, or excessively corrective responses that reduce self-confidence.

### 3.2.3. LLMs as Mediators

The mediator role involves bridging gaps in access, interpretation, or communication. Examples include systems like DoNotPay, which guides users through legal processes [12], and AI translation tools that facilitate multilingual interaction in real time [16]. Mediation is particularly relevant in contexts where individuals face complex or opaque systems, whether legal, bureaucratic, or linguistic [4].

The defining requirement for mediators is interpretability. The system must not only provide content but also explain reasoning in a manner that is accessible to the user [6]. Reducing jargon, simplifying complex processes, and highlighting actionable steps are crucial [5]. Risks include liability for incorrect advice and the danger of oversimplification, where critical nuances are lost in the pursuit of accessibility [12], [16].

### 3.2.4. LLMs as Curators

The curator role is oriented toward information filtering, synthesis, and explanation. Tools such as Perplexity AI, Consensus, and Elicit embody this function by aggregating data from multiple sources and presenting it in an intelligible form [14], [15]. Unlike companions or coaches, curators do not seek to foster long-term relational bonds. Instead, they create clarity and trust in contexts where information is overwhelming [6].

Curators must provide trustworthy sourcing, verifiable citations, and a balance between breadth and depth of coverage. The primary risks involve hallucinated references, overconfidence in uncertain findings, and the challenge of ensuring that simplified summaries preserve the complexity of original evidence [7], [14].

### 3.3. Visualization of the Framework

The four roles can be positioned along two intersecting dimensions: emotional versus cognitive orientation, and personal versus collective use cases [4]. Companions and coaches align more closely with personal contexts, where the user's growth and well-being are the central focus. Mediators and curators are more collective, since they often support interactions between the user and broader systems such as law, bureaucracy, or scientific knowledge. Similarly, companions and mediators emphasize relational and emotional aspects, while coaches and curators focus on cognitive growth and decision-making [2].

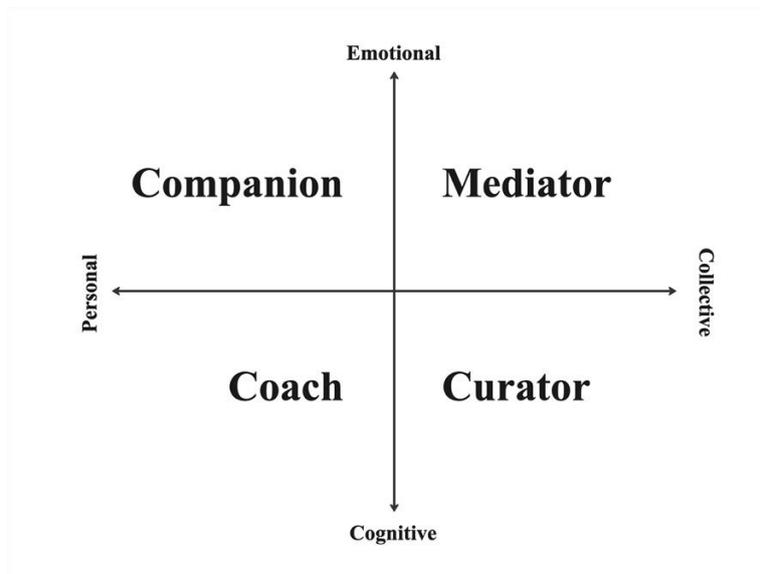

Figure 1. Quadrant Framework of LLM Roles

This visualization highlights that the roles are not mutually exclusive but can overlap. For example, an educational tutor might function both as a coach (providing structured feedback) and as a curator (summarizing information) [10], [14]. Similarly, an elder-care companion might integrate mediation functions when assisting with healthcare navigation [8].

### 3.4. Core Design Principles Across Roles

Although each role has unique requirements, several cross-cutting design principles emerge [5], [6]:
1. **Transparency and Explainability.** Users must understand why an LLM provided a given recommendation or response. This is particularly crucial in mediator and curator roles [6].
2. **Personalization and Cultural Sensitivity.** LLMs should adapt their tone, examples, and explanations to align with user backgrounds [10], [11].
3. **Safety Nets and Guardrails.** Systems should prevent harmful outputs, whether by filtering unsafe language, redirecting users to professional resources, or clarifying limitations of the AI [4].
4. **Memory and Continuity.** Effective support requires sustained interaction. Memory mechanisms can track progress, recall past interactions, and maintain consistency, balanced with privacy protections [6].
5. **Balancing Empathy and Reliability.** Particularly in companion roles, it is possible for systems to be overly empathetic without ensuring factual correctness [17].

### 3.5. Intersections Between Roles

While the framework distinguishes between companions, coaches, mediators, and curators, in practice these roles frequently overlap. For example, a tutoring system may function simultaneously as a coach, guiding learners through practice, and as a curator, summarizing external resources to support knowledge acquisition [10], [14]. Similarly, elder-care companions may incorporate mediation functions when assisting users with healthcare navigation [8]. Such overlaps create design trade-offs. A system that acts as both a companion and a coach must balance empathy with accuracy: excessive encouragement may obscure errors, while strict correction may undermine relational trust. Mediators and curators face parallel tensions, since simplification for accessibility may conflict with the need for comprehensive citation. Future work should investigate systematic methods for role integration, including adaptive switching mechanisms, multi-role evaluation benchmarks, and guidelines for mitigating conflicts when roles converge.

### 3.6. Summary

This framework repositions LLMs from transactional tools to relational systems that assume distinct roles in human-centered support. By categorizing them as companions, coaches, mediators, and curators [4], it creates a vocabulary for evaluating how design choices affect their impact on users. At the same time, the boundaries between roles are not always clear-cut. In practice, systems often blend functions, such as a tutor that acts both as a coach and a curator [10], [14], or an elder-care companion that also serves as a mediator [8]. These overlaps generate design trade-offs, where empathy must be balanced with accuracy, and simplification must not come at the cost of verifiability. Recognizing these intersections is essential for building evaluation methods that account for multi-role dynamics. The next section illustrates these roles with comparative case studies, highlighting both practical successes and design challenges across diverse domains [8], [10], [12], [14].

## 4. COMPARATIVE CASE STUDIES

This section examines real-world examples of LLM-based systems through the lens of the framework introduced above. Each case study illustrates how the roles of companion, coach, mediator, and curator manifest in practice, the benefits they provide, and the challenges they encounter. Together, these case studies demonstrate both the promise and complexity of deploying LLMs as human-centered support systems [4], [6].

Table 1. Comparative Roles of LLMs in Human-Centered Support Systems

| Role | Example Systems | Key Design Principles | Main Risks |
|---|---|---|---|
| Companion | Replika, Character.AI, ElliQ | Emotional sensitivity, personalization, continuity | Over-reliance, privacy concerns, commercialization |
| Coach | Duolingo Max, Khanmigo | Adaptive scaffolding, feedback transparency, pacing | Inaccurate feedback, discouragement, equity gaps |
| Mediator | DoNotPay, AI translation tools | Interpretability, simplification, accessibility | Liability, oversimplification, mistranslation |
| Curator | Perplexity, Consensus, Elicit | Citation accuracy, verifiability, balanced synthesis | Hallucinations, biased sources, overconfidence |

### 4.1. Companion: Emotional and Social Support

One of the most widely discussed examples of LLMs functioning as companions is Replika, a conversational AI marketed as an artificial friend [8], [9]. Users engage with Replika to alleviate loneliness, discuss emotions, or simply practice casual conversation. Similarly, platforms like Character.AI allow users to interact with customizable characters, while ElliQ integrates conversational AI into a physical robot designed to support older adults [8].

The benefits of companion systems are significant. For elderly individuals living alone, conversational agents can reduce isolation and encourage adherence to healthy routines [8]. For younger users, they can provide a safe space to practice social skills or manage anxiety [9]. Studies report that users often feel comfort and reduced stress when engaging with such systems [8], [9], highlighting the power of perceived presence even when users are aware of the artificial nature.

The challenges are equally pronounced. Emotional sensitivity is difficult to encode, and failures can damage user trust [17]. There is also a risk of over-dependence, where users substitute artificial companionship for human relationships [8]. Ethical concerns arise around commercialization, as users may become financially or emotionally invested in premium features [9].

### 4.2. Coach: Guided Skill-Building

The role of coach is best exemplified by educational and training platforms that integrate LLMs into structured learning experiences. Duolingo Max employs GPT-4 to create role-play scenarios that allow users to practice conversations in a foreign language. It also offers an "explain my answer" feature, which provides learners with personalized feedback when they make mistakes [10]. Khanmigo, developed by Khan Academy, adopts a similar approach by guiding students through math and science problems with Socratic-style questioning [11].

Beyond education, coaching paradigms have also emerged in other sensitive domains. For instance, researchers have documented how LLM-based systems can support mental health coaching, offering conversational scaffolding and reflective prompts that supplement traditional therapeutic approaches [15]. Similarly, early work has evaluated LLMs as training mediators, providing structured dispute-resolution scenarios that simulate real-world negotiations [16]. These applications demonstrate that the coaching role extends beyond academics, with potential to shape user growth in professional, psychological, and interpersonal contexts.

The primary benefit of LLMs as coaches is scalability. Personalized instruction, traditionally limited to one-on-one tutoring, becomes accessible to millions of learners at once. The adaptive nature of LLMs allows for real-time tailoring of lessons, explanations, and feedback, which can sustain learner motivation. Furthermore, the conversational interface makes practice more engaging compared to static exercises.

Yet the risks are considerable. Accuracy in subject matter remains a persistent challenge, especially in domains where precision is critical. An incorrect explanation in mathematics or science may mislead students and create long-term misconceptions. In addition, coaching systems must balance correction with encouragement; overly critical feedback may discourage learners, while insufficient correction may fail to promote growth. Finally, educational equity is at stake: while these tools expand access, they may also widen gaps if advanced features are locked behind paid subscriptions. The success of LLM coaches therefore depends on aligning pedagogy, accessibility, and technical reliability.

### 4.3. Mediator: Bridging Gaps in Access

The mediator role is demonstrated by systems designed to make complex or inaccessible domains more navigable. DoNotPay, often described as the "robot lawyer," automates tasks such as contesting parking tickets or filing consumer complaints. By translating legal jargon into plain language and generating ready-to-use templates, it aims to democratize access to justice [12]. Beyond law, LLM-powered translation systems such as DeepL Chat and Google's live translation in Meet serve as mediators across languages, enabling real-time communication between speakers of different linguistic backgrounds [16].

The benefits of LLMs as mediators are clear. They reduce barriers for individuals who would otherwise lack access to expensive legal services or professional translation. They also save time and reduce cognitive load by presenting information in accessible forms [4]. This role has particular relevance in addressing global inequalities, where marginalized communities are disproportionately excluded from legal and bureaucratic processes [12].

The risks, however, are substantial. Incorrect legal advice could have serious consequences for users, raising questions about liability and consumer protection [12]. Similarly, mistranslations in professional or diplomatic settings could damage relationships or create misunderstandings [16]. Moreover, there is a danger of oversimplification: while simplifying complex processes is valuable, it may inadvertently obscure critical nuances that users need to make informed decisions [6]. Mediators therefore require especially strong guardrails, including clear disclaimers, explainable reasoning, and in many cases, human oversight [5].

### 4.4. Curator: Information Filtering and Explanation

The curator role is embodied by platforms that help users navigate the overwhelming abundance of information available online. *Perplexity AI* acts as a conversational search engine that provides concise answers with citations [14], while *Consensus* specializes in synthesizing evidence from peer-reviewed scientific literature. *Elicit* further extends this role by supporting systematic reviews, enabling researchers to rapidly extract insights from large bodies of academic work [15]. The benefits of curator systems are most evident in contexts where the volume of information exceeds individual capacity for review [14], [15]. By filtering, summarizing, and structuring knowledge, these systems allow users to make decisions with greater confidence and efficiency. For students, researchers, and professionals, curators reduce the time required to locate relevant materials and highlight connections that might otherwise remain hidden [6].

Challenges center on trust and verifiability. Users may place undue confidence in summaries that contain hallucinated citations or misrepresentations of source material [7], [14]. Striking the right balance between breadth and depth is difficult: summaries that are too broad may lack actionable detail, while those that are too narrow may miss important perspectives. To fulfill their potential, curator systems must integrate rigorous source validation, transparent citation practices, and mechanisms that allow users to trace back to original materials [5], [6].

### 4.5. Cross-Domain Insights

Across these four case studies, several patterns emerge. Companions prioritize emotional sensitivity and continuity, but they risk over-dependence [8], [9]. Coaches emphasize adaptive feedback and sustained growth, yet they must carefully balance accuracy and encouragement [10], [11]. Mediators expand access by reducing complexity, but their outputs must be both interpretable and reliable [12], [16]. Curators filter vast information sources, but they depend on strong verification mechanisms to build trust [14], [15].

These insights reinforce the value of conceptualizing LLMs not as monolithic tools but as systems that assume distinct roles in relation to human users [4]. By articulating these roles, we can better evaluate the design choices, opportunities, and risks that shape their deployment in sensitive contexts [6], [17].

## 5. EVALUATION AND METRICS

Evaluating large language models in human-centered support systems requires a shift from conventional benchmarks that emphasize accuracy, speed, or computational efficiency [6]. While such metrics remain important, they are insufficient for assessing systems that function as companions, coaches, mediators, or curators [4]. In these contexts, the central question is not only whether the system produces the correct output but also whether it promotes trust, growth, well-being, and informed decision-making [13], [17]. This section proposes an evaluation framework that combines functional performance with human-centered outcomes.

### 5.1. Beyond Accuracy

Accuracy has long been the standard measure of performance in natural language processing, but its limitations become clear in relational roles [7]. A companion system may provide factually accurate information yet fail to convey empathy, leaving the user unsatisfied [8]. Similarly, a coach may offer a correct answer but discourage learning if feedback is framed in a demotivating way [10], [11]. For mediators and curators, accuracy alone does not guarantee usefulness, since users also require clarity, accessibility, and actionable insight [12], [14], [15].

Therefore, evaluations must extend beyond factual correctness to encompass qualities such as emotional resonance, engagement, and user confidence [13]. For example, a companion should be assessed on its ability to sustain positive emotional states [9], a coach on whether learners demonstrate improved skills over time [10], a mediator on whether users can complete tasks more effectively [12], and a curator on whether decisions are made with greater clarity and reduced cognitive burden [14], [15].

### 5.2. Trust and Transparency Metrics

Trust is foundational for any human-centered support system [6], [17]. Users must believe that the system operates reliably, that its limitations are visible, and that its reasoning can be understood [5]. For LLMs, this involves transparency about data sources, reasoning steps, and the boundaries of model capability [13].

Metrics for trust may include:

- **Perceived reliability:** measured through user surveys or longitudinal studies that capture whether individuals feel confident relying on the system [13].
- **Interpretability:** assessed by evaluating whether users can understand explanations of outputs, such as why a particular legal recommendation was generated [12].
- **Consistency:** measured by tracking whether the system produces stable responses over repeated interactions with similar prompts [6].

In companion roles, trust may be expressed in emotional safety [8], [9], while in mediator and curator roles, it may manifest in the ability to verify claims or trace outputs to external references [14], [15].

### 5.3. Longitudinal Measures of Growth and Engagement

Since human-centered roles involve ongoing support, evaluation must consider longitudinal outcomes [17]. A coaching system should not only answer questions correctly but also lead to measurable improvements in learning outcomes over time [10], [11]. Similarly, a companion should sustain engagement without fostering unhealthy dependency [8], [9].

Relevant longitudinal measures include:

- **Engagement metrics:** duration and frequency of use across time periods, indicating whether users sustain interaction or abandon the system [13].
- **Progress metrics:** improvement in skill mastery, test performance, or confidence levels among learners [10], [11].
- **Emotional well-being metrics:** self-reported scales of stress, loneliness, or satisfaction when interacting with companion systems [9].
- **Dependency metrics:** indicators of over-reliance, such as reduced willingness to seek alternative sources of support [8].

These measures allow designers to distinguish between systems that provide temporary novelty and those that produce sustained positive outcomes [17].

### 5.4. Domain-Specific Benchmarks

Each role requires domain-specific benchmarks tailored to its unique purpose [5], [6], [13]:

- **Companion:** Evaluations may draw from psychology, using validated scales such as the UCLA Loneliness Scale or the Positive and Negative Affect Schedule (PANAS) to measure emotional impact [8], [9].
- **Coach:** Benchmarks include academic performance tests, retention of knowledge, and measures of learner motivation. In language learning, pronunciation accuracy or conversation fluency can serve as quantifiable outcomes [10], [11].
- **Mediator:** Metrics include task completion rates, error reduction in legal or bureaucratic processes, and user comprehension scores when interpreting simplified explanations [12], [16].
- **Curator:** Benchmarks focus on citation accuracy, coverage of relevant sources, and user ability to make well-informed decisions after consulting summaries [14], [15].

These domain-specific metrics complement general measures of accuracy and trust, providing a holistic assessment of effectiveness [13], [17].

### 5.5. Toward a Unified Evaluation Framework

Bringing these elements together, a unified framework for evaluating LLMs in human-centered roles should integrate three levels of assessment [6], [13], [17]:

1. **Functional metrics:** accuracy, latency, stability, and computational efficiency [7].
2. **Human-centered metrics:** trust, interpretability, emotional resonance, and user confidence [5], [6].
3. **Longitudinal metrics:** sustained engagement, skill progression, emotional well-being, and avoidance of over-dependence [8], [9], [10].

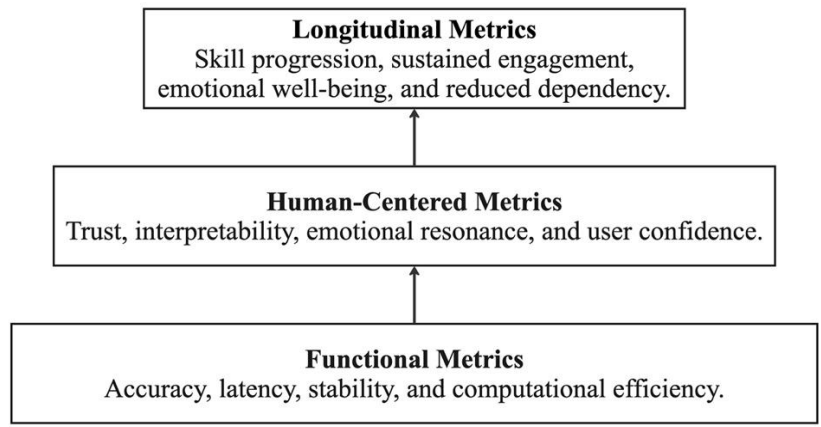

Figure 2. Evaluation Framework for LLMs in Support Systems

This layered approach ensures that systems are not judged solely on technical grounds but also on their ability to foster meaningful, safe, and sustainable forms of human support [4], [17].

### 5.6. Summary

Evaluation of LLMs as human-centered support systems requires moving beyond accuracy to incorporate trust, emotional resonance, skill development, and long-term user outcomes [6], [13]. By aligning metrics with the specific demands of the companion, coach, mediator, and curator roles, researchers and practitioners can better understand both the benefits and risks of these systems [8], [10], [12], [14]. A unified framework that integrates functional, human-centered, and longitudinal measures provides a foundation for responsible design and deployment [17].

## 6. RISKS, CHALLENGES, AND ETHICAL CONCERNS

The promise of LLMs as human-centered support systems is tempered by significant risks. Because these models increasingly interact with users in sensitive contexts—health, education, law, and emotional support—the potential for harm extends beyond technical error into the social and ethical domains [4], [6], [7]. This section outlines five categories of concern: over-reliance and dependency, misinformation and hallucination, bias and personalization risks, privacy and data security, and economic as well as accessibility challenges [13], [17].

### 6.1. Over-Reliance and Dependency

A central risk of companion and coaching systems is that users may become overly dependent on artificial support. For example, studies of Replika users suggest that some individuals perceive their AI companions as replacements for human relationships [8], [9]. This raises concerns about social isolation and reduced real-world interaction [17]. In coaching contexts, learners may rely on AI feedback rather than cultivating independent problem-solving skills [10], [11]. While support is valuable, over-reliance can reduce resilience, critical thinking, and willingness to seek human guidance [2].

Designers must therefore implement safeguards that encourage balanced use [5]. Companion systems can include reminders to engage with real-world social networks, while coaching platforms might deliberately withhold direct answers to foster persistence [11]. Evaluation should monitor not only engagement levels but also patterns that suggest unhealthy dependency [13].

## 6.2. Misinformation and Hallucination

Hallucination, which occurs when a model generates confident but incorrect responses, poses particular danger in mediator and curator roles [7]. In legal contexts, fabricated case citations could mislead users into presenting invalid arguments [12]. In healthcare, inaccurate medical summaries could influence treatment decisions with serious consequences [8]. Even in education, repeated exposure to subtle inaccuracies may result in cumulative misunderstandings [10].

Addressing this risk requires multiple strategies. Technical approaches include grounding responses in verifiable sources and using retrieval-augmented generation to reduce hallucination rates [15]. Equally important are user-facing strategies: systems should communicate uncertainty, present citations, and explicitly mark the boundaries of their reliability [5], [6]. Without such measures, the persuasive fluency of LLMs can mask the fragility of their knowledge [7], [14].

## 6.3. Bias and Personalization Risks

Like all data-driven systems, LLMs inherit the biases present in their training corpora [7]. When functioning as companions, biases in language use may affect how empathy is expressed across cultures and demographics [8], [9]. As coaches, biased feedback could disadvantage learners from underrepresented backgrounds [10], [11]. As mediators, models risk reinforcing systemic inequities if they simplify legal or bureaucratic processes in ways that obscure structural disadvantages [12].

Personalization, while often beneficial, amplifies these concerns [6]. Systems that tailor responses to user data may inadvertently reinforce stereotypes or echo-chamber effects. For example, a curator that consistently surfaces sources aligned with a user's prior beliefs may reduce exposure to alternative perspectives [14], [15]. Mitigating these risks requires diverse training datasets, fairness audits, and mechanisms that deliberately introduce perspective-taking into AI-generated outputs [4], [13].

## 6.4. Privacy and Data Security

Human-centered roles often require continuity and personalization, which in turn demand the storage of sensitive user data [6]. Companions may retain intimate details of emotional states [8], [9]; coaches may track learning progress and mistakes [10]; and mediators may process legal or financial information [12]. These data streams raise significant privacy concerns, particularly when stored by commercial platforms [4].

Without strict safeguards, there is a risk of unauthorized access, misuse, or commodification of user data [13]. Furthermore, systems with long-term memory may create new vulnerabilities if users are not fully aware of what information is being stored [17]. Solutions include transparent consent mechanisms, encryption, local data storage options, and the ability for users to selectively delete memory [5], [6]. Privacy cannot be treated as an afterthought; it is a prerequisite for trust in human-centered support systems [17].

## 6.5. Economic and Accessibility Challenges

While many LLM systems are marketed as democratizing access, economic realities complicate this narrative [4]. Premium subscriptions often gate advanced features such as personalized feedback or long-term memory [10], [11]. In effect, those who can afford to pay receive superior forms of support, while others are restricted to more limited experiences [13]. This risks exacerbating existing inequalities in education, healthcare, and access to justice [12].

Accessibility also extends beyond cost. Companion systems that do not adapt to diverse linguistic and cultural contexts may exclude marginalized groups [8]. Coaching systems designed for one educational standard may not translate to others [11]. True human-centered design must therefore include not only affordable pricing models but also inclusive approaches that account for global diversity [4], [6].

## 6.6. Risk Matrix

Not all risks carry the same severity or urgency. Table 2 presents a risk matrix that categorizes concerns by likelihood and impact. High-severity risks include misinformation and hallucination [7], [12], privacy and data security violations [6], [17], and over-reliance leading to social or cognitive dependency [8], [9]. Medium-severity risks include cultural bias, personalization errors, and accessibility gaps [10], [11]. Lower-severity but still relevant risks include economic inequalities created by subscription models [4]. Distinguishing risks by severity helps prioritize mitigation strategies: technical safeguards such as retrieval-augmented generation are most critical for misinformation, while regulatory frameworks are vital for liability and privacy protections.

Table 2. Risk Matrix: High, Medium, Low severity with examples and mitigation strategies

| Risk Category | Severity | Examples | Mitigation Strategy |
| --- | --- | --- | --- |
| Misinformation & Hallucination | High | Fabricated legal citations, incorrect medical summaries [7], [12] | Retrieval grounding, uncertainty displays, expert oversight |
| Privacy & Data Security | High | Retention of sensitive conversations, unauthorized data use [6], [17] | Encryption, local storage, consent management, regulatory standards |
| Over-Reliance & Dependency | High | Substituting AI companions for human relationships [8], [9] | Usage reminders, balanced feedback, dependency monitoring |
| Bias & Personalization Risks | Medium | Reinforcing stereotypes, echo chambers [10], [11] | Fairness audits, diverse training data, perspective-taking |
| Accessibility & Equity Gaps | Medium | Paywalled features, lack of cultural adaptation [4], [11] | Inclusive design, tiered pricing models, multilingual support |
| Economic Inequality | Low | Advanced functions gated by premium subscriptions [10] | Subsidized access, open-source alternatives |

**6.7. Summary**

The risks outlined above highlight the complexity of deploying LLMs in roles that extend into human-centered support [6], [13], [17]. Over-reliance may weaken resilience [8], [9], hallucinations may cause harm [7], [12], biases may reinforce inequities [4], [10], privacy violations may erode trust [6], [17], and economic barriers may deepen divides [11], [13]. Addressing these challenges requires a combination of technical innovation, ethical design, and regulatory oversight [4], [5]. The following section outlines future directions that can mitigate these risks while harnessing the potential of LLMs to serve as companions, coaches, mediators, and curators in a safe and responsible manner.

**7. FUTURE DIRECTIONS**

The risks and challenges identified in the previous section underscore the importance of carefully guiding the development of LLMs as human-centered support systems [4], [6], [7]. While technical innovation continues at a rapid pace, the long-term impact of these systems will depend on how well they are designed, evaluated, and governed [13], [17]. This section identifies five future directions: unified evaluation frameworks, hybrid human–AI systems, memory and privacy innovations, cross-domain benchmarking, and policy as well as governance strategies.

**7.1. Unified Evaluation Frameworks**

A major limitation of current practice is the reliance on fragmented evaluation methods [13]. Education systems focus on learning outcomes [10], [11], healthcare systems emphasize clinical safety [8], [9], and information retrieval tools measure citation accuracy [14], [15]. A unified evaluation framework that integrates functional, human-centered, and longitudinal metrics would provide a more holistic basis for comparison [6], [17]. Such frameworks should assess not only accuracy and efficiency but also trust, emotional well-being, and long-term user growth [5], [13]. This would allow stakeholders to make informed decisions about when and how to deploy LLMs in sensitive support contexts [4], [17].

**7.2. Hybrid Human–AI Support Systems**

Future deployments will likely move toward hybrid models in which LLMs act as co-pilots rather than replacements [16]. For example, an AI mediator might draft legal documents that are then reviewed by a human lawyer [12], or an AI coach might provide daily practice exercises while a teacher oversees long-term progress [10]. Hybrid approaches leverage the scalability of AI while retaining the accountability and ethical judgment of human professionals [6]. Research is needed to determine optimal divisions of labor between human and AI agents, as well as strategies for seamless handoff when complex or high-risk situations arise [17].

**7.3. Memory and Privacy Innovations**

Continuity of interaction is central to effective companionship, coaching, and mediation, yet persistent memory raises privacy concerns [6], [17]. Future research should explore architectures that allow selective memory retention, user-controlled deletion, and encrypted local storage [5]. Innovations in federated learning and privacy-preserving computation may enable systems to personalize experiences without exposing sensitive data to centralized servers [4]. Establishing clear standards for memory transparency, ensuring that users always know what is remembered and why, will be essential for sustaining trust [13], [17].

### 7.4. Cross-Domain Benchmarking

Most evaluations today remain confined to single domains, limiting the ability to generalize findings [13]. Cross-domain benchmarking would allow researchers to assess how LLM roles transfer across contexts [6]. For instance, techniques that improve transparency in legal mediation might also enhance trust in healthcare companions [12]. Comparative studies could identify universal design principles while clarifying where domain-specific adaptations are necessary [10], [11]. Such benchmarking would also provide a foundation for regulating LLMs in sensitive environments by setting minimum performance thresholds across diverse applications [4], [17].

### 7.5. Policy and Governance

Finally, the societal impact of LLMs requires thoughtful policy frameworks [4]. Regulations should address issues such as liability for errors in mediator systems [12], informed consent for memory in companion systems [8], [9], and equitable access to educational coaching systems [10], [11]. Governance may also involve certification standards, whereby systems are evaluated against ethical benchmarks before deployment in high-stakes contexts [6], [17]. Policymakers, researchers, and developers must collaborate to ensure that innovation is balanced with accountability [5]. Without governance, commercial incentives alone may drive practices that prioritize profit over well-being [7].

### 7.6. Summary

Future directions for LLMs as human-centered support systems must go beyond technical advancement [6], [7], [13]. Unified evaluation frameworks can provide clarity [13], [17], hybrid models can preserve human accountability [16], memory innovations can protect privacy [5], [6], [17], cross-domain benchmarking can reveal transferable design principles [10], [11], [12], and policy can ensure equitable and ethical deployment [4], [6], [7]. Together, these strategies provide a roadmap for transforming LLMs into sustainable, trustworthy partners that augment human capacity without eroding autonomy or dignity [17].

## 8. CONCLUSION

Large language models are no longer confined to the role of conversational engines that provide answers on demand [6], [7]. As they become integrated into applications that touch on education, healthcare, legal support, companionship, and knowledge management, their function increasingly aligns with what can be described as human-centered support systems [4], [13], [17]. This paper has argued that to fully understand and responsibly deploy these technologies, evaluation must move beyond technical accuracy alone and instead examine the relational roles they assume in human lives [5], [6].

The framework introduced here identifies four such roles: companions, coaches, mediators, and curators [4]. Each role brings unique opportunities and distinct challenges. Companions can alleviate loneliness and provide emotional presence but risk encouraging over-reliance [8], [9]. Coaches can democratize access to personalized skill-building but require careful calibration to ensure accuracy and encouragement [10], [11]. Mediators can open pathways to justice, healthcare, or communication, yet they carry liability when errors occur [12], [16]. Curators can help users navigate overwhelming volumes of information, but their value depends on transparency and verifiability [14], [15].

Through comparative analysis of real-world examples, this paper has shown that these roles require tailored design principles, including transparency, personalization, safety mechanisms, continuity, and a balance between empathy and reliability [5], [6], [17]. It has also argued for evaluation frameworks that integrate functional, human-centered, and longitudinal measures, ensuring that these systems are judged not only by their outputs but also by their impact on growth, trust, and well-being [13], [17].

At the same time, the risks of over-reliance, misinformation, bias, privacy violations, and unequal access highlight the importance of cautious deployment [7], [8], [9], [12], [13]. Future directions such as unified evaluation frameworks [13], hybrid human–AI support systems [16], privacy innovations [5], [6], [17], cross-domain benchmarking [10], [11], and regulatory governance [4] offer a roadmap for mitigating these risks while preserving the benefits.

Ultimately, the question is not whether LLMs will play a role in human-centered support, but how this role will be shaped [6], [17]. If guided responsibly, these systems can extend access to care, education, and knowledge, fostering resilience and empowerment [4], [10]. If left unchecked, they may exacerbate inequities or create new forms of dependency [7], [13]. Researchers, practitioners, and policymakers share a responsibility to ensure that LLMs evolve as tools that accompany, guide, and empower, rather than manipulate or diminish, the humans who rely on them [17].

This study is limited by its conceptual methodology. The framework synthesizes case studies and prior literature but does not include a formal empirical evaluation. As such, its conclusions should be viewed as provisional and subject to validation. Furthermore, the examples analyzed focus primarily on well-documented systems in education, legal aid, companionship, and knowledge discovery; the framework may require adaptation in other domains such as creative industries, workplace collaboration, or non-Western cultural contexts. Finally, while the framework identifies design principles and risks, it does not prescribe technical implementation details. These limitations underscore the need for future empirical research, including user studies, comparative evaluations, and cross-cultural analyses, to validate, refine, and extend the roles and principles outlined here.

## AUTHOR

**Zhiyin Zhou** is an MS in Information Systems candidate at Trine University with a background in product design (Syracuse University) and design management (Pratt Institute). Her research focuses on generative AI in clinical research, healthcare, and human-centered digital knowledge systems. She is also a technical project manager at CRIO (Clinical Research IO), leading the development of eClinical platforms.

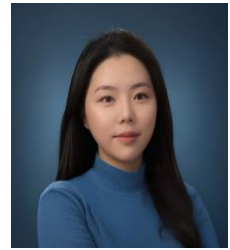